# Altruism and anxiety: Engagement with online community support initiatives (OCSIs) during Covid-19 lockdown in the UK and Ireland

*Covid-19 online community support initiatives*

Camilla Elphick (camilla.elphick@open.ac.uk)[1]*; Avelie Stuart (a.stuart@exeter.ac.uk)[2]; Richard Philpot (r.philpot@lancaster.ac.uk)[3]; Zoe Walkington (z.walkington@open.ac.uk)[1]; Lara Frumkin (lara.frumkin@open.ac.uk)[1]; Min Zhang (min.zhang@open.ac.uk)[1]; Mark Levine (mark.levine@lancaster.ac.uk)[3]; Blaine Price (b.a.price@open.ac.uk)[1]; Graham Pike (graham.pike@open.ac.uk)[1]; Bashar Nuseibeh (bashar.nuseibeh@open.ac.uk)[1&4]; Arosha Bandara (arosha.bandara@open.ac.uk)[1]

[1]The Open University, Walton Hall, Kents Hill, Milton Keynes, MK7 6AA
[2]University of Exeter, Stocker Road, Exeter, EX4 4PY
[3]Lancaster University, Bailrigg, Lancaster, LA1 4YW
[4]Lero - The Irish Software Research Centre, University of Limerick, Limerick, Ireland

Given concerns about mental health during periods of Covid-19 lockdown, it important to understand how engagement with online Covid-19 related material can affect mood. In the UK and Ireland, online community support initiatives (OCSIs) have emerged to help people manage their lives. Yet, little is known about how people engaged with these or whether they influenced subsequent mood. We conducted surveys to explore how people in the UK and Ireland engaged with OCSIs, and found that 70% did so to offer support (e.g. to provide company). Those who did so reported feeling significantly calmer afterwards, those who engaged for general concerns (e.g. in response to anti-social behaviour) reported feeling significantly more anxious afterwards, but there was no difference in reported mood for those who engaged for other reasons (e.g. to share experiences or views). Thus, engaging with an OCSI for altruistic purposes might help to make people feel calmer.

**Data availability statement:** The data that support the findings of this study are available on request from the corresponding author.

**Acknowledgements:** This work was supported by the Citizen Forensics project, funded by the UK EPSRC (EP/R033862/1 and EP/R013144/1), and Science Foundation Ireland (SFI 13/RC/2094). The funders had no role in the design of the paper, analysis, decision to publish, or preparation of the manuscript.

* Corresponding author, e-mail: Camilla Elphick, camilla.elphick@open.ac.uk

On 11th March, 2020, the WHO declared Covid-19 a pandemic. By May 29th (the time of writing), the SARS-CoV-2 coronavirus had infected at least 5,816,706 and killed at least 360,437 people worldwide (Johns Hopkins University, n.d.). In an attempt to slow the spread, many countries introduced 'lockdown' measures, where citizens were asked to stay at home, including the UK (on March 23rd) and Ireland (on March 27th). Both countries remained in lockdown throughout April.

As a result, citizens were forced, overnight, to practice 'social distancing'. This inevitably created some issues (e.g. how to 'shield' the vulnerable – van Bunnik et al., 2020) and amplified others (e.g. the risk of domestic abuse – Usher et al., 2020), but it also inspired a variety of online community support initiatives (OCSIs), designed to help people manage difficulties faced during Covid-19. For instance, Covid Mutual Aid (n.d.) was launched to connect support groups with vulnerable people.

Given the uncertainty generated by Covid-19, as well as the consequences of social distancing (e.g. to jobs, relationships), there were also concerns about mental health. However, research shows that online activity can help. As well as existing online mental health mechanisms, such as Woebot (n.d.), which alleviates depression and anxiety (Fitzpatrick, Darcy, & Vierhile, 2017), Ferrara and Yang (2015) found that engagement with positive content on social media can develop positive emotions. They showed that despite no in-person interaction on Twitter between a person who tweets and a person who engages with the tweet, positive, negative, or neutral tweets were related to sentiment expressed in the previous tweet. Social networks are also associated with spreading happiness, which can have health benefits (Fowler & Christakis, 2008).

There are also emotional contagion effects in the other direction (Ferrara & Yang, 2015; Fowler & Christakis, 2008). For instance, Coviello et al. (2014) found that when people post emotionally depressed messages when it is raining, the subsequent messages of friends living in places where it is not raining are also more emotionally depressed. In other words, an emotion expressed via social media may spread, despite the reason for the emotion being absent in the recipient of the message.

OCSIs have been set up with altruism and community spirit in mind, and research suggests that altruism is linked with happiness and well-being (see Post, 2005). Therefore, given the literature on the positive spread of feelings online, engagement with altruism (online) during Covid-19 could help to generate positive feelings. Conversely, there are valid concerns that online engagement could amplify Covid-19 catastrophising, which can stem from feeling powerless and increase feelings of anxiety (see Wiederhold, 2020).

Therefore, the aim of the current research was to ask whether engaging with OCSIs generated a positive or negative mood in those who engaged with them (focusing on anxiety, empowerment, and reassurance), and whether or not different reasons for engaging might play different roles in subsequent mood. Anxiety was chosen as previous research demonstrates a link between it and disease outbreaks (e.g. Wheaton et al. 2012), and anxiety and empowerment are related to catastrophising (see Wiederhold, 2020). Reassurance was chosen as OCSIs evolved to support individuals at a time of lockdown, so we wanted to explore whether they did indeed reassure those who engaged with them for support.

## Methods

### Participants

Participants were recruited from the first day of lockdown in the UK (n = 261), between 23rd March, 2020 and 30th April 2020, and between 30th March and 30th April for participants from IRL (n = 65), resulting in a final sample of 326. The 314 that provided their age were between 18 and 80 ($M = 43.58$, $SD = 13.89$). 255 identified as female, 56 as male, and 2 as non-binary. The remaining 13 did not respond. 172 described themselves as

Caucasian, 6 as Asian, 4 as Mixed race, 3 as Black, and 1 as Latin and 1 as Mediterranean. The remaining 139 did not respond.

**Materials and Procedure**

Participants were invited to take part in a survey that was built and run using Qualtrics (n.d.). They were recruited via the research group website (https://citizenforensics.org/study), survey circle (https://www.surveycircle.com) and via social media, using official university accounts and those of the researchers. After providing informed consent, participants completed the survey, after which they were debriefed. Participants that did not complete the survey or indicated that they were under 18 were excluded from the final sample, resulting in a 71% completion rate.

**Design**

Data were analysed using a principal components exploratory factor analysis, with promax rotations, to explore why people engaged with an OCSI after the WHO declared a pandemic (11th March). Factor analyses allow one to understand how responses to different items cluster into fewer categories (factors). The results were then used to determine whether different motivations to engage were related to subsequent mood, using regression.

The three different mood types measured were anxiety; empowerment; and reassurance. These mood types were measured using participants' retrospective impressions of their mood after engaging with an OCSI. Participants rated their feelings for each as either much less (-2); slightly less (-1); slightly more (1); or much more (2).

## Results

**Why People Engaged with an OCSI**

To explore how people engaged with OCSIs and whether this was related to subsequent mood, we conducted an exploratory (principal components) factor analysis (promax rotation) on 25 questions concerning participants' engagement with OCSIs that were designed after pilot work exploring online communication by the authors of the current study. 150 participants from the sample had used an OCSI (46%) and provided responses that were used in the analysis. One item failed to load at above .4 (*To journal or blog*), so this item was removed and the analysis run again. Then one item double loaded at above .4 (*To get or give feedback*), so this item was removed and the analysis run again.

The Kaiser-Mayer-Olkin measure verified the sampling adequacy for the analysis, KMO = .78, and all but one of the KMO values for individual items were greater than .58, which is above the acceptable limit of .50 (Field, 2013) (the value of this item was .49). Seven factors had eigenvalues over Kaiser's criterion of 1 and in combination explained 59.58% of the variance. The scree plot justified retaining seven factors.

Table 1 shows the OCSI engagement factors loading after rotation. This indicates that the items cluster into the following factors: factor 1 *offering support*; factor 2 *communication*; factor 3 *action and awareness*; factor 4 *company and safety*; factor 5 *networking and problem solving*; factor 6 *general concerns*; and factor 7 represents being *directly affected* by Covid-19. Most people engaged with an OCSI to offer support (70%), followed by action and awareness (57%), communication (56%), networking and problem solving (41%), general concerns (40%), because they were directly affected (21%), and for company or safety (6%). Thus, people overwhelmingly used OCSIs for altruistic reasons. The threshold of what is considered acceptable reliability (using Cronbach's *a*) varies between studies (see Taber, 2018) and inspection of Table 1 confirmed that the items loaded into logical factors. So, we used the model as a guide for exploring whether reasons for engaging with OCSIs were related to subsequent mood.

**Table 1**

*Summary of exploratory Factor Analysis results for reasons for engaging with an OCSI*

|  | Offering support | Communication | Action & awareness | Company & safety | Networking | General concern | Directly affected |
|---|---|---|---|---|---|---|---|
| Support others who felt unsafe | **0.794** | -0.155 | -0.123 | -0.095 | -0.043 | 0.208 | -0.032 |
| Help someone feel less anxious | **0.772** | 0.169 | -0.028 | 0.009 | -0.059 | 0.03 | 0.208 |
| Give company | **0.648** | -0.013 | -0.238 | 0.169 | 0.071 | 0.175 | -0.055 |
| Get or give help | **0.547** | -0.011 | 0.37 | -0.05 | 0.063 | -0.24 | 0.001 |
| Be proactive | **0.502** | 0.221 | 0.256 | -0.112 | 0.033 | -0.212 | -0.172 |
| Share experiences or views | 0.023 | **0.789** | 0.092 | -0.059 | -0.205 | 0.119 | -0.221 |
| Connect with or build communities | 0.048 | **0.699** | -0.29 | 0.074 | 0.291 | -0.063 | 0.021 |
| Communicate informally | -0.088 | **0.686** | -0.089 | 0.19 | 0.227 | 0.09 | 0.008 |
| Be reactive | 0.091 | -0.112 | **0.722** | 0.177 | 0.099 | -0.057 | -0.053 |
| For advice/info | -0.368 | -0.114 | **0.701** | 0.059 | -0.06 | 0.005 | 0.222 |
| Gain perspective | 0.01 | 0.299 | **0.556** | -0.108 | -0.116 | 0.178 | 0.17 |
| Report something | 0.171 | -0.22 | **0.551** | 0.115 | 0.249 | 0.242 | -0.173 |
| For company | 0.15 | 0.109 | 0.036 | **0.742** | -0.168 | -0.007 | 0.171 |
| Make friends or network | -0.139 | 0.096 | 0.096 | **0.737** | 0.323 | -0.156 | -0.138 |
| I felt unsafe | -0.074 | -0.044 | 0.121 | **0.67** | -0.02 | 0.281 | -0.049 |
| Solve problems | 0.071 | -0.016 | -0.034 | 0.122 | **0.761** | -0.002 | 0.067 |
| Collaborate | -0.072 | 0.337 | 0.12 | -0.049 | **0.568** | -0.04 | 0.153 |
| Advocate or raise awareness | -0.125 | 0.242 | 0.133 | -0.248 | **0.446** | 0.305 | -0.055 |
| Anti-social behaviour | -0.017 | 0.141 | 0.056 | 0.169 | -0.267 | **0.692** | -0.095 |
| Give advice/info | 0.059 | 0.153 | 0.003 | -0.084 | 0.271 | **0.552** | 0.044 |
| Affected someone else | 0.322 | -0.164 | -0.046 | -0.016 | 0.219 | **0.512** | 0.125 |
| Affected me or someone I live with | -0.059 | -0.225 | 0.019 | -0.095 | 0.187 | -0.015 | **0.83** |
| Feel less anxious | 0.132 | 0.121 | 0.201 | 0.141 | -0.057 | 0.016 | **0.665** |
| Eigenvalues | 5.16 | 1.87 | 1.86 | 1.36 | 1.24 | 1.12 | 1.11 |
| % of variance | 22.44 | 8.13 | 8.08 | 5.89 | 5.38 | 4.95 | 4.81 |
| *a* | 0.74 | 0.61 | 0.57 | 0.65 | 0.57 | 0.48 | 0.41 |

**Relationships Between OCSI Engagement and Subsequent Mood**

First, we asked participants how they felt after engaging with an OCSI and 155 responded. When combining scores for anxiety, empowerment, and reassurance, to create an overall mood score (range -6 and +6), the mean score showed that participants were in better overall moods after engaging with an OCSI than before ($M = 3.10$, $SD = 1.49$).

We therefore conducted a second (principal components) factor analysis (promax rotation), to explore whether the three mood types (anxiety; empowerment; and reassurance) loaded to form an overall mood score that could be used as an outcome variable in the multiple regression. We found that while empowerment and reassurance loaded into a single factor at above .4 (.49 and .51 respectively), anxiety did not (.29). Therefore, we calculated the mean scores for *empowerment and reassurance* to form one outcome variable, and retained *anxiety* as a second outcome variable.

Finally, we conducted two exploratory multiple linear regressions, using the enter method, with all OCSI engagement factors, to see which predicted anxiety, or empowerment and reassurance. The first model significantly predicted *anxiety*, $F(7,148) = 2.16$, $p = .04$, $R^2 = .10$, $R^2_{Adjusted} = .05$. Offering support, $t(148) = 2.72$, $p = .01$, (Beta = .68), CI [0.19,1.17], and general concerns, $t(148) = 2.36$, $p = .02$, (Beta = -.68), CI [-1.26 -0.11] were significantly associated with anxiety. The second regression model with OSCI factors predicting *empowerment and reassurance* was not significant $F(7,149) = 1.08$, $p = .38$ and was not interpreted.

**Table 2**

*Multiple linear regression analysis of the OSCI factors predicting anxiety*

| Factor | Variables | B | $SE_B$ | 95% CI | ß | p | t | F | $R^2$ |
|---|---|---|---|---|---|---|---|---|---|
| | (Intercept) | 0.76*** | 0.11 | [0.54, 0.98] | | <.001 | 6.70 | 2.16* | 0.10 |
| 1 | Offering support | 0.68** | 0.25 | [0.19, 1.17] | 0.26 | .007 | 2.72 | | |
| 2 | Communication | 0.12 | 0.23 | [-0.34, 0.59] | 0.05 | .60 | .52 | | |
| 3 | Action and awareness | -0.14 | 0.30 | [-0.74, 0.46] | -0.04 | .64 | -.46 | | |
| 4 | Company and safety | -0.31 | 0.38 | [-1.07, 0.44] | -0.07 | .42 | -.82 | | |
| 5 | Networking and problem solving | 0.22 | 0.28 | [-0.33, 0.78] | 0.08 | .43 | .79 | | |
| 6 | General concerns | -0.68* | 0.29 | [-1.26, -0.11] | -0.22 | .02 | -2.36 | | |
| 7 | Directly affected | 0.15 | 0.36 | [–0.57, 0.86] | 0.03 | .69 | .41 | | |

*Note.* N = 149. ß = standardized regression coefficient. 95% CI = 95 percent confidence intervals.

*p < .05, **p < .01, ***p < .001.

Thus, after engaging with an OCSI to offer support, participants reported subsequently feeling calmer than if they engaged for other reasons, but when they engaged for general concerns, they reported feeling more anxious.

## Discussion

The present research explored the ways that people engaged with OCSIs to help manage the challenges of Covid-19, and relationships between engagement and subsequent mood. Engagement was categorised into seven factors: *offering support*; *communication*; *action and awareness*; *company and safety*; *networking and problem solving*; *general concerns*, and being *directly affected* by Covid-19. Seventy percent of people responded that they had engaged with OCSIs to offer support. We also explored whether different motivations for engaging with an OCSI would affect feelings of anxiety, empowerment or reassurance in different ways.

We started with the relationship between OCSI engagement and subsequent anxiety, and found that participants that had engaged to offer support (such as providing company) were calmer afterwards. This suggests that, despite the seriousness of Covid-19, it is possible to improve mood by engaging with positive Covid-19-related online activities, and supports the notion that altruism is good for the altruist (Post, 2005). Our finding that offering support may reduce anxiety is also in line with meta-analytic evidence suggesting that volunteering has a positive impact on mental health (Jenkinson et al., 2013). In contrast, the finding that those engaged in general concerns (such as responding to anti-social behaviour) became more anxious, supports the notion of Covid-19 catastrophising, which Wiederhold (2020) describes as the amplification of anxiety stemming from powerlessness. These results suggest that it may not be online engagement with Covid-19 itself that shapes mood, but whether the engagement is altruistic or not. However, the effects described were only seen when it came to feelings of anxiety (or calm), rather than feelings of empowerment or reassurance.

When it came to empowerment, the non-significant findings were a surprise, as several OCSI engagement items appeared to be empowering (e.g. to connect with or build communities; to solve problems; or to advocate or raise awareness). This might be because empowerment is a *predictor* of mood (anxiety) when it comes to catastrophising, rather than an *outcome* of OCSI engagement. The findings for reassurance could be explained by sample size, as only 27% of the participants engaged to seek support (e.g. because they felt unsafe or they needed company). It might also be that combining empowerment and reassurance into one outcome variable was confounded by the different motivations to engage in the first place. However, it is also worth noting that we measured mood using participants' retrospective impressions, and the model only explained 10 percent of the variance, so it would be worth testing relationships between OCSIs and mood also via other means.

In conclusion, as it is important to know the impact of Covid-19 related online activities on mental health, further work is needed to measure Covid-19 related mood more effectively as people engage these activities; to explore whether Covid-19 related mood spreads more widely online and offline; and whether Covid-19 related mood is a precursor to mental health, with a more representative sample. Nevertheless, this exploratory research suggests that despite the challenges of the lockdown stages of a pandemic, altruistic engagement with an OCSI may help people feel calmer.

## References

Covid Mutual Aid (n.d.), retrieved, 23rd May, 2020, from https://covidmutualaid.org

Coviello, L., Sohn, Y., Kramer, A. D. I., Marlow, C., Franceschetti, M., Christakis, A., & Fowler, J. H. (2014) Detecting emotional contagion in massive social networks. *PLoS ONE, 9*(3): e90315. doi:10.1371/journal.pone.0090315

Ferrara, E., & Yang, Z. (2015) Measuring emotional contagion in social media. *PLoS ONE, 10*(11): e0142390. doi:10.1371/journal.pone.0142390

Field, A. (2013). *Discovering statistics using IBM SPSS statistics.* Sage Publications.

Fitzpatrick, K. K., Darcy, A., & Vierhile, M. (2017). Delivering cognitive behavior therapy to young adults with symptoms of depression and anxiety using a fully automated conversational agent (Woebot): A randomized controlled trial. *JMIR Mental Health*, *4*(2), e19. https://doi.org/10.2196/mental.7785

Fowler, J. H., & Christakis, N. A. (2008). Dynamic spread of happiness in a large social network: longitudinal analysis over 20 years in the Framingham Heart Study. *Bmj*, *337*, a2338. https://doi.org/10.1136/bmj.a2338

Jenkinson, C. E., Dickens, A. P., Jones, K., Thompson-Coon, J., Taylor, R. S., Rogers, M., … Richards, S. H. (2013). Is volunteering a public health intervention? A systematic review and meta-analysis of the health and survival of volunteers. *BMC Public Health, 13*(1), 773. https://doi.org/10.1186/1471-2458-13-773

Johns Hopkins University Covid-19 dashboard (n.d.), retrieved, 23rd May, 2020, from https://coronavirus.jhu.edu/map.html

Ring (n.d.), retrieved, 23rd May, 2020, from https://en-uk.ring.com

Post, S. G. (2005). Altruism, happiness, and health: It's good to be good. *International Journal of Behavioral Medicine*, *12*(2), 66-77. https://doi.org/10.1207/s15327558ijbm1202_4

Taber, K.S. (2018) The use of Cronbach's Alpha when developing and reporting research instruments in science education. *Res Sci Educ* 48, 1273–1296. https://doi.org/10.1007/s11165-016-9602-2

Usher, K., Bhullar, N., Durkin, J., Gyamfi, N., & Jackson, D. (2020). Family violence and COVID-19: Increased vulnerability and reduced options for support. *International Journal of Mental Health Nursing, [advanced online publication]*. https://doi.org/10.1111/inm.12735

van Bunnik, B. A., Morgan, A. L., Bessell, P., Calder-Gerver, G., Zhang, F., Haynes, S., ... & Lepper, H. C. (2020). Segmentation and shielding of the most vulnerable members of the population as elements of an exit strategy from COVID-19 lockdown. M*edRxiv*. [preprint]. https://doi.org/10.1101/2020.05.04.20090597

Wheaton, M.G., Abramowitz, J.S., Berman, N.C., Fabricant, L.E. & Ulatunji, B.O. (2012). Psychological predictors of anxiety in response to the H1N1 (swine flu) pandemic. *Cognitive Therapy and Research*, 36(3), 210-218. DOI: 10.1007/s10608-011-9353-3

Wiederhold, B. K. (2020). Social media use during social distancing. *Cyberpsychology, Behavior, and Social Networking, 2*3(5), 275-276.